\documentclass[aps,prd,twocolumn,nofootinbib,showpacs]{revtex4}

\usepackage{graphicx,epsfig}

\begin{document}

\title{Application of the Principle of Maximum Conformality to the Top-Quark Charge Asymmetry at the LHC}

\author{Sheng-Quan Wang}
\email[email:]{sqwang@cqu.edu.cn}

\author{Xing-Gang Wu}
\email[email:]{wuxg@cqu.edu.cn}
\address{Department of Physics, Chongqing University, Chongqing 401331, P.R. China}
\address{State Key Laboratory of Theoretical Physics, Institute of Theoretical Physics, Chinese Academy of Sciences, Beijing 100190, P.R. China}

\author{Zong-Guo Si}
\email[email:]{zgsi@sdu.edu.cn}
\address{Department of Physics, Shandong University, Jinan, Shandong 250100, P.R. China}

\author{Stanley J. Brodsky}
\email[email:]{sjbth@slac.stanford.edu}
\address{SLAC National Accelerator Laboratory, Stanford University, Stanford, California 94039, USA}

\date{\today}

\begin{abstract}

The Principle of Maximum Conformality (PMC) provides a systematic and process-independent method to derive renormalization scheme- and scale- independent fixed-order pQCD predictions. In Ref.\cite{pmc3}, we studied the top-quark charge asymmetry at the Tevatron. By applying the PMC, we have shown that the large discrepancies for the top-quark charge asymmetry between the Standard Model estimate and the CDF and D0 data are greatly reduced. In the present paper, with the help of the Bernreuther-Si program, we present a detailed PMC analysis on the top-quark pair production up to next-to-next-to-leading order level at the LHC. After applying PMC scale setting, the pQCD prediction for the top-quark charge asymmetry at the LHC has very small scale uncertainty; e.g., $A_{\rm C}|_{\rm 7 TeV;PMC} =\left(1.15^{+0.01}_{-0.03}\right)\%$, $A_{\rm C}|_{\rm 8 TeV;PMC} =\left(1.03^{+0.01}_{+0.00}\right)\%$, and $A_{\rm C}|_{\rm 14 TeV;PMC} =\left(0.62^{+0.00}_{-0.02}\right)\%$. The corresponding predictions using conventional scale setting are: $A_{\rm C}|_{\rm 7 TeV;Conv.} =\left(1.23^{+0.14}_{-0.14}\right)\%$, $A_{\rm C}|_{\rm 8 TeV;Conv.} =\left(1.11^{+0.17}_{-0.13}\right)\%$, and $A_{\rm C}|_{\rm 14 TeV;Conv.} =\left(0.67^{+0.05}_{-0.05}\right)\%$. In these predictions, the scale errors are predicted by varying the initial renormalization and factorization scales in the ranges $\mu^{\rm init}_r\in[m_t/2,2m_t]$ and $\mu_f\in[m_t/2,2m_t]$. The PMC predictions are also in better agreement with the available ATLAS and CMS data. In addition, we have calculated the top-quark charge asymmetry assuming several typical cuts on the top-pair invariant mass $M_{t\bar{t}}$.  For example, assuming $M_{t\bar{t}}>0.5 ~ {\rm TeV}$ and $\mu_f=\mu^{\rm init}_r =m_t$, we obtain $A_{\rm C}|_{\rm 7 TeV;PMC}=2.67\%$, $A_{\rm C}|_{\rm 8 TeV;PMC}=2.39\%$, and $A_{\rm C}|_{\rm 14 TeV;PMC}=1.28\%$.


\end{abstract}

\maketitle

\section{Introduction}

The hadroproduction of the top quark plays a crucial role in testing the Standard Model (SM) as well as for searches for new physics. The properties of the top quark, such as its mass, production cross sections, decay rates, and its charge asymmetries, have been measured at both the Tevatron and the Large Hadron Collider (LHC). The experimental data are generally compatible with the SM predictions; however, the predictions for the top quark forward-backward asymmetry in $p \bar{p} \to t \bar{t} X$ at the Tevatron are in substantial disagreement with the experimental measurements~\cite{asym1,asym2,asym3,asym4, hpAfb,jgAfb, wsAfb, elew1, elew2,elew3,elew4,elew5,elew6}. In fact, if one uses conventional scale setting, i.e., guessing the renormalization scale and its range, the predicted $t\bar{t}$ forward-backward asymmetry deviates significantly from the Tevatron CDF and D0 measurements~\cite{CDF1,CDF2,D01,D02}. The difference between theory and experiment ranges up to a $3.4~\sigma$ standard deviation for a $t\bar{t}$ invariant mass $M_{t\bar{t}}>450~{\rm GeV}$~\cite{tev3s}. It is clearly important to understand the origin of this discrepancy -- is it new physics or an artifact of the calculational method ?

We have shown that the large discrepancies for the top-quark charge asymmetry with pQCD predictions observed at the Tevatron can be attributed to an improper choice of the renormalization scale~\cite{pmc3}. In the conventional procedure, the renormalization scale $\mu_r$ is fixed at the value $\mu_r^{\rm init}$, which is usually chosen as $m_t$ in order to eliminate the large logarithmic terms $\ln^k m_t^2/(\mu^{\rm init}_r)^2$. It should be emphasized that this procedure for setting the renormalization scale and its range is only a guess; in fact, the resulting predictions are scheme dependent, violating renormalization group invariance. In the case of its QED analog, the $\mu^+ \mu^-$ charge asymmetry in $e^+ e^- \to \mu^+ \mu^- X$, this method disagrees with the Standard Gell Mann-Low method for scale setting, where the renormalization scale is set by the photon virtuality~\cite{gml}; in fact, a new renormalization scale and effective number of leptons $n_\ell$ appears at each order of perturbation theory. The conventional procedure of guessing the renormalization scale and its range in pQCD results in an unnecessary systematic error for $t\bar{t}$-pair production, and it can even lead to incorrect finite-order predictions.

Renormalization group invariance implies that the prediction for a physical observable cannot depend on the choice of the initial renormalization scale~\cite{rgi1,rgi2,rgi3,pmc5,pmc12} or the choice of the renormalization scheme. The Principle of Maximum Conformality (PMC) provides a systematic and unambiguous way to set the renormalization scale and to eliminate the renormalization scheme and scale uncertainty for {\it fixed-order pQCD predictions}~\cite{pmc1,pmc2,pmc4, pmc6,pmc8,pmc9,pmc11}.

The running behavior of the QCD coupling constant is governed by the $\beta$-function of its renormalization group equation. The guiding principle of the PMC is that all terms proportional to the QCD $\beta$-functions $\beta_0$, $\beta_1$, $\beta_2$, $\ldots$ should be resummed into the running coupling; this procedure determines the correct renormalization scale and the effective number of quark flavors at each perturbative order. The resulting pQCD series then has the same coefficients of the $\beta=0$ ``conformal" series which is renormalization-scheme independent. In the $N_C \to 0$ Abelian limit~\cite{Brodsky:1997jk}, this procedure agrees with Gell Mann-Low scale setting. One can also use the PMC to derive ``commensurate scale relations"~\cite{csr1} such as the ``Generalized Crewther Relation"~\cite{gcr1,gcr2,gcr3} which relate observables to each other independent of the choice of renormalization scheme.

After applying the PMC, we obtain the optimal scale of the process at each order in pQCD, and the resulting theoretical predictions are essentially free of initial scale dependence. Furthermore, the divergent renormalon series do not appear in the PMC prediction, and the pQCD convergence is generally greatly improved.

The PMC provides the underlying principle for the well-known Brodsky-Lepage-Mackenzie scheme~\cite{blm}, and it is applicable at all orders in pQCD. Some recent higher order PMC applications can be found in Refs.~\cite{jpsi,pom,higbbgg,higrr,chamm,rho:fact, Z:decay, bctocham, btopi,gluon}. In particular, we have shown that after applying the PMC, the SM predictions for the top-quark charge asymmetry at the Tevatron have only $1\sigma$ deviation from the CDF and D0 measurements~\cite{pmc3}; the large discrepancies of the top-quark charge asymmetry between the SM estimate and the data are thus greatly reduced.

The top-quark charge asymmetry at the LHC for the $p p \to t \bar t X$ process is defined as
\begin{eqnarray}
A_{\rm C}&=&\frac{N(\Delta |y|>0)-N(\Delta |y|<0)} {N(\Delta |y|>0)+N(\Delta |y|<0)},  \label{asyc}
\end{eqnarray}
where $\Delta |y|=|y_{t}|-|y_{\bar{t}}|$ is the difference between the absolute rapidity of the top and anti-top quarks, and $N$ is the number of events. Measurements of the top-quark charge asymmetry at the LHC have been reported in Refs.~\cite{CMS1,CMS2,CMS3,ATLAS1,ATLAS2}. The recent preliminary ATLAS+CMS measurements give $A_{\rm C}|_{\rm 7 TeV}=(0.5\pm0.7\pm0.6)\%$~\cite{ATLAS3}.

In contrast to the Tevatron $ p \bar p \to t \bar t X$ processes, the asymmetric channel $q\bar{q}\to t\bar{t}$ provides a small pQCD contribution to the top-pair production at the LHC, and the symmetric channel $gg\to t\bar{t}$ provides the dominant contribution. Thus, the predicted charge asymmetry at the LHC is usually smaller than the one at the Tevatron. Two typical SM predictions for the charge asymmetry at the LHC are: $A_{\rm C}|_{\rm 7 TeV} =(1.15\pm0.06)\%$ and $A_{\rm C}|_{\rm 8 TeV}=(1.02\pm0.05)\%$ for Ref.~\cite{jgAfb}; $A_{\rm C}|_{\rm 7 TeV} =(1.23\pm0.05)\%$ and $A_{\rm C}|_{\rm 8 TeV}=(1.11\pm0.04)\%$ for Ref.~\cite{wsAfb}. The uncertainties of those two SM predictions are the scale errors obtained by using the conventional renormalization scale and range, $\mu_{r} \in[m_t/2,2m_t]$, and by fixing the factorization scale $\mu_{f} \equiv \mu_{r}$. Thus if one uses conventional scale setting, the resulting scale uncertainties provide the dominant error for the pQCD prediction.

In this paper we shall apply the PMC to predict the top-quark charge asymmetry at the LHC based on the analysis of the top-pair hadroproduction up to next-to-next-to-leading order (NNLO) level. It is found that a more accurate top-quark asymmetry with less scale uncertainties can be achieved. We shall show that the PMC predictions are in agreement with the available ATLAS and CMS data within errors; since the renormalization scale uncertainties are essentially eliminated, the constraints on new beyond the Standard Model physics are considerably strengthened.

The remaining sections of the paper are organized as follows. In Sec.~\ref{sect2}, we present the calculational technology for applying PMC scale setting to the top-quark charge asymmetries at the LHC. The Bernreuther-Si (BS) program~\cite{wsAfb} for doing the NNLO QCD corrections, together with the electroweak corrections, are adopted for our present purposes. We then present the numerical results and discussions in Sec.~\ref{sect3}. A summary is given in Sec.~\ref{sect4}.

\section{The top-quark charge asymmetry using PMC scale setting}
\label{sect2}

We applied PMC scale setting to determine the renormalization scales for the top-pair hadroproduction cross sections at the Tevatron in Refs.~\cite{pmc3,pmc4}. For self-consistency, we shall present the main formulas here; interested readers may turn to Refs.~\cite{pmc3,pmc4} for a detailed analysis. We shall then apply the same technology to deal with the top-quark charge asymmetry at the LHC.

Total hadronic cross section for the top-quark pair production, $H_1 H_2 \to {t\bar{t} X}$, can be obtained from the convolution of the factorized partonic cross-section $\hat \sigma_{ij}$ with the parton luminosities ${\cal L}_{ij}$
\begin{equation}
\sigma = \sum_{i,j} \int\limits_{4m^2_{t}}^{S}\, ds \,\, {\cal L}_{ij}(s, S, \mu_f) \hat \sigma_{ij}(s,\alpha_s(\mu_r),\mu_r,\mu_f) ,
\end{equation}
with the parton luminosity
\begin{displaymath}
{\cal L}_{ij} = {1\over S} \int\limits_s^S {d\hat{s}\over \hat{s}} f_{i/H_1}\left(x_1,\mu_f\right) f_{j/H_2}\left(x_2,\mu_f\right),
\end{displaymath}
where $x_1= {\hat{s} / S}$ and $x_2= {s / \hat{s}}$. Here $S$ denotes the hadronic center-of-mass (CM) energy squared and $s=x_1 x_2 S$ is the subprocess center-of-mass energy squared. The functions $f_{i/H_{1,2}}$ are the parton distribution functions (PDFs), and $\hat \sigma_{ij}$ is the partonic subprocess cross section, where $(ij) = \{(q{\bar q}), (gg), (gq), (g\bar{q})\}$ stands for the four relevant production channels.

Writing the numerator and the denominator of the asymmetry $A_{C}$ in powers of $\alpha_s$, we obtain
\begin{eqnarray}
A_{C} &=& \frac{\alpha_s^{3} N_{1}+\alpha_s^{4} N_{2}+ {\cal O}(\alpha_s^5)}{\alpha_s^{2} D_{0} +\alpha_s^{3} D_{1}+\alpha_s^{4} D_{2} +{\cal O}(\alpha_s^5)} \nonumber\\
&=& \frac{\alpha_s}{D_{0}}\left[N_{1}+ \alpha_s\left(N_{2}-\frac{D_{1} N_{1}}{D_{0}}\right) + \right.\nonumber\\
&& \quad\quad\left. \alpha_s^2 \left(\frac{D_1^2 N_1}{D_0^2} -\frac{D_1 N_2}{D_0} -\frac{D_2 N_1}{D_0}\right) +\cdots \right],
\end{eqnarray}
where the $D_i$-terms stand for the total cross-sections at certain $\alpha_s$-order and the $N_i$-terms stand for the asymmetric cross-sections at certain $\alpha_s$-order.

We will refer to $A^{\rm (BS)}_{\rm C}$ as the asymmetry under the conventional scale setting by using the BS-program~\cite{wsAfb}. When one uses conventional scale setting, the $N_{1}D_{1}/D_{0}$ term and the $N_{2}$ term have the same importance. Since the NNLO $N_{2}$ term is not available at present, one then has to use the lowest-order $\mathcal{O}(\alpha_s^{2})$ cross-section in the denominator and the $\mathcal{O}(\alpha_s^{3})$ term in the numerator; i.e., one can only address the ``LO asymmetry", $A^{\rm (BS)}_{C}=\frac{N_{1}}{D_{0}} \alpha_s$. In contrast, we will show that, after PMC scale setting, the size of the NNLO corrections for both the total cross-sections and the asymmetry are decreased by about one order of magnitude. Thus in the case of the PMC, the NNLO-terms $N_2$ and $D_2$ can be safely neglected, and we obtain:
\begin{displaymath}
A^{\rm (PMC)}_{C} =\frac{N_{1}}{D_{0}}\alpha_s \left[1-\alpha_s\left(\frac{D_{1}}{D_{0}}\right) +\alpha_s^2 \left(\frac{D_1^2}{D_0^2}\right) \right].
\end{displaymath}
This result can be resummed to a convenient form; i.e., the ``NLO-asymmetry"~\cite{pmc3}:
\begin{equation}
A^{\rm (PMC)}_{C}= \frac{\alpha_s^{3} N_{1}}{\alpha_s^{2} D_{0} +\alpha_s^{3} D_{1}} \;.
\end{equation}
When one includes the $\mathcal{O}(\alpha^2_s\alpha)$ and $\mathcal{O}(\alpha^2)$ electroweak contributions, the top-pair asymmetry can be written as~\cite{pmc3}
\begin{eqnarray}
A^{\rm (PMC)}_{\rm C}={\alpha^3_s N_1+\alpha^2_s \alpha \tilde{N}_1+\alpha^2\tilde{N}_0 \over \alpha^2_{s} D_0 + \alpha^3_{s} D_1}, \label{pmcac}
\end{eqnarray}
where $\tilde{N}_1$ and $\tilde{N}_0$ terms stand for the QCD-electroweak and the pure electroweak asymmetric cross sections, respectively. These terms can be obtained from Refs.\cite{ttnlo1,ttnlo2,ttnlo3,ttnlo4,elew1, elew2,elew3,elew4,elew5,elew6, spine1,spine2,spine3, spine4,spine5, ttnnlo1,ttnnlo2,moch1, moch2,beneke2, hathor}, or they can be numerically calculated using the BS program~\cite{wsAfb}.

In order to compare the PMC prediction with the asymmetry assuming conventional scale setting $A^{\rm (BS)}_{\rm C}$, we will rewrite the PMC asymmetry $A_{\rm C}^{\rm (PMC)}$ as~\cite{pmc3}
\begin{widetext}
\begin{eqnarray}
A_{\rm C}^{\rm (PMC)}&=&  \left\{\frac{\sigma^{\rm BS,LO}_{\rm tot}} {\sigma^{\rm PMC,NLO}_{\rm tot}} \right\} \left\{ \frac{{\overline{\alpha}_s}^3\left(\overline{\mu}^{\rm PMC, NLO}_r\right)} {{\alpha^{\rm BS}_s}^3 \left(\mu^{\rm conv}_r\right)} \left. A^{\rm (BS)}_{\rm C}\right|_{\alpha^3_s} + \frac{{\overline{\alpha}_s}^2\left(\overline{\mu}^{\rm PMC, NLO}_r\right)} {{\alpha^{\rm BS}_s}^2 \left(\mu^{\rm conv}_r\right)} \left. A^{\rm (BS)}_{\rm C}\right|_{\alpha^2_s \alpha}+ \left. A^{\rm (BS)}_{\rm C}\right|_{\alpha^2} \right\}, \label{pmcasy}
\end{eqnarray}
\end{widetext}
where the symbol ``BS" stands for the prediction calculated by using the BS-program using conventional scale setting, and ``PMC" stands for the corresponding value after applying the PMC. The $A^{\rm (BS)}_{\rm C}|_{\alpha^3_s}$, $A^{\rm (BS)}_{\rm C}|_{\alpha^2_s \alpha}$, and $A^{\rm (BS)}_{\rm C}|_{\alpha^2}$ stand for the predicted QCD, the QCD-electroweak, and the pure electroweak asymmetry, respectively. The $\alpha^{\rm BS}_s \left(\mu^{\rm conv}_r\right)$ is the coupling constant assuming conventional scale setting. In addition, we have defined an effective coupling constant ${\overline{\alpha}_s} \left( \overline{\mu}^{\rm PMC, NLO}_r \right)$ for the asymmetric part of the QCD contributions, which is the weighted average of the strong coupling constant for the asymmetric $(q\bar{q})$-channel~\cite{pmc9}; i.e., in using the effective coupling constant ${\overline{\alpha}_s} \left(\overline{\mu}^{\rm PMC, NLO}_r \right)$, one obtains the same $(q\bar{q})$-channel NLO cross section as that obtained from ${\alpha}_s \left(\mu^{\rm PMC, NLO}_r \right)$. \\

\section{Numerical results} \label{sect3}

\subsection{Top-pair total cross-section at the LHC}

We take the top quark mass $m_{t}=173.1$ GeV and the PDF as CTEQ6.6M~\cite{pdf66}. We shall use the BS program~\cite{wsAfb} to do our calculation, and as a cross check, we also adopt the HATHOR program~\cite{hathor} to calculate the total cross-sections. Taking the same input parameters in these two programs, we obtain the same results for the top-pair total cross sections up to NNLO level.

\begin{widetext}
\begin{center}
\begin{table}[htb]
\begin{tabular}{|c||c|c|c|c||c|c|c|c|}
\hline
& \multicolumn{4}{c||}{Conventional scale setting} & \multicolumn{4}{c|}{PMC scale setting} \\
\hline
~~~ ~~~    &~~~LO~~~  &~~~NLO~~~  &~~~NNLO~~~ &~~~ {\it Total} ~~~&~~~LO~~~  &~~~NLO~~~  &~~~NNLO~~~ &~~~ {\it Total} ~~~\\
\hline
$(q\bar{q})$-channel & 22.645 & 3.302 & 1.798 & 27.773 & 21.612 & 6.959 & -0.728 & 27.690 \\
\hline
$(gg)$-channel   & 77.431 & 45.171 & 10.473 & 133.070 & 77.140 & 52.708  & 8.463  & 140.188 \\
\hline
$(gq)$-channel    & 0.000 & -0.412 & 1.380 & 1.024 & 0.000 & -0.412 & 1.380  & 1.024 \\
\hline
$(g\bar{q})$-channel & 0.000 & -0.411 & 0.232& -0.182 & 0.000 & -0.411 & 0.232  & -0.182 \\
\hline
sum     & 100.076  & 47.650 & 13.883 & 161.686 & 98.752 & 58.844  & 9.346  & 168.720 \\
\hline
\end{tabular}
\caption{The top-pair production cross sections (in unit: pb) at the LHC assuming conventional and PMC scale settings, respectively. Each of the four production channels, $(q\bar{q})$-channel, $(gg)$-channel, $(gq)$-channel, and $(g\bar{q})$-channel, are calculated separately. The CM collision energy is assumed to be $\sqrt{S}=7$ TeV, and $\mu^{\rm init}_r=\mu_f=m_t$.}  \label{tab2}
\end{table}

\begin{table}[htb]
\begin{tabular}{|c||c|c|c|c||c|c|c|c|}
\hline
& \multicolumn{4}{c||}{Conventional scale setting} & \multicolumn{4}{c|}{PMC scale setting} \\
\hline
~~~ ~~~    &~~~LO~~~  &~~~NLO~~~  &~~~NNLO~~~ &~~~ {\it Total} ~~~&~~~LO~~~  &~~~NLO~~~  &~~~NNLO~~~ &~~~ {\it Total} ~~~\\
\hline
$(q\bar{q})$-channel & 28.915 & 4.050 & 2.230 & 35.233 & 27.534 & 8.737 &-0.993 & 35.087\\
\hline
$(gg)$-channel   & 113.741 & 65.071 & 14.503 & 193.316 & 113.315 & 75.776 & 11.543 & 203.256 \\
\hline
$(gq)$-channel    & 0.000 & 0.181 & 1.983 & 2.141 & 0.000 & 0.181 & 1.983 & 2.141 \\
\hline
$(g\bar{q})$-channel & 0.000 & -0.496 & 0.362 & -0.133 & 0.000 & -0.496 & 0.362 & -0.133\\
\hline
sum     & 142.656 & 68.806 & 19.078 & 230.557 & 140.849 & 84.198 & 12.895 & 240.351 \\
\hline
\end{tabular}
\caption{The top-pair production cross sections (in unit: pb) at the LHC assuming conventional and PMC scale settings, respectively. Each of the four production channels, $(q\bar{q})$-channel, $(gg)$-channel, $(gq)$-channel, and $(g\bar{q})$-channel, are calculated separately. The CM collision energy is assumed to be $\sqrt{S}=8$ TeV, and $\mu^{\rm init}_r=\mu_f=m_t$. }\label{tab8}
\end{table}

\begin{table}[ht]
\begin{tabular}{|c||c|c|c|c||c|c|c|c|}
\hline
& \multicolumn{4}{c||}{Conventional scale setting} & \multicolumn{4}{c|}{PMC scale setting} \\
\hline
~~~ ~~~    &~~~LO~~~  &~~~NLO~~~  &~~~NNLO~~~ &~~~ {\it Total} ~~~&~~~LO~~~  &~~~NLO~~~  &~~~NNLO~~~ &~~~ {\it Total} ~~~\\
\hline
$(q\bar{q})$-channel & 71.977 & 8.896 & 5.083 & 86.048 & 68.014 & 20.743 & -2.892 & 85.416 \\
\hline
$(gg)$-channel   & 479.744 & 259.009 & 49.377 & 788.203 & 477.974 & 299.007 & 36.927 & 823.185 \\
\hline
$(gq)$-channel    & 0.000 & 9.058 & 7.563 & 16.874 & 0.000 & 9.058 & 7.563 & 16.874 \\
\hline
$(g\bar{q})$-channel & 0.000 & 0.053 & 1.889 & 1.880 & 0.000 & 0.053 & 1.889 & 1.880 \\
\hline
sum     & 551.721 & 277.016 & 63.912 & 893.006 & 545.988 & 328.861 & 43.487 & 927.356 \\
\hline
\end{tabular}
\caption{The top-pair production cross sections (in unit: pb) at the LHC assuming conventional and PMC scale settings, respectively. Each of the four production channels, $(q\bar{q})$-channel, $(gg)$-channel, $(gq)$-channel, and $(g\bar{q})$-channel, are calculated separately. The CM collision energy is assumed to be $\sqrt{S}=14$ TeV, and $\mu^{\rm init}_r=\mu_f=m_t$. } \label{tab14}
\end{table}
\end{center}
\end{widetext}

We present the numerical results before and after PMC scale setting at the LHC with the collision energies $\sqrt{S}=7$ TeV, $8$ TeV, and 14 TeV in Tables \ref{tab2}, \ref{tab8}, and \ref{tab14}, respectively. Note that the results listed in the {\it Total}-column are not the simple addition of the corresponding LO, NLO and NNLO cross sections, since they are obtained using the Sommerfeld re-scattering formula to treat the Coulomb contributions~\cite{pmc3}. From these Tables, we observe:

\begin{itemize}
\item At the LHC, the symmetric $(gg)$-channel provides the dominant contribution to the total top-pair production cross section. The $(q\bar{q})$-channel, the $(gq)$-channel, and the $(g\bar{q})$-channel are asymmetric, among which the $(q\bar{q})$-channel dominantly determines the charge asymmetry. At the LO level, the $(q\bar{q})$-channel does not discriminate between the final top quark and top-antiquark, so their distributions are symmetric. At the NLO level and higher orders, either virtual or real gluon emission will cause differences between the distributions of the top quark and antiquark production, thus leading to an observable top-quark charge asymmetry.

\item At the LHC, the total cross section for each channel increases with increasing CM collision energy $\sqrt{S}$, and the total cross section is dominated by the symmetric $(gg)$-channel. This can be compared with the Tevatron case, in which the asymmetric $(q\bar{q})$-channel provides the dominant contribution to the total cross section. Thus at the LHC the charge asymmetry shall be highly diluted by the $(gg)$-channel, and a smaller charge asymmetry is expected at the LHC compared to the Tevatron.

\item After applying PMC scale setting, the pQCD convergence has been greatly improved. For example, for the $(q\bar{q})$-channel with $\sqrt{S}=7$ TeV, the ratio for the total cross section at the NNLO level and the NLO level, $|\sigma_{q\bar{q}}^{\rm NNLO}/\sigma_{q\bar{q}}^{\rm NLO}|$, is about $54\%$ when using the conventional scale setting. Thus in order to derive a consistent asymmetry up to NNLO, one must consider the asymmetric contribution from the NNLO $(q\bar{q})$-channel, which may be sizable. In contrast, this ratio reduces to $\sim 10\%$ after applying the PMC. This shows that after PMC scale setting, the change to the asymmetry from the NNLO term is greatly suppressed, being consistent with the discussions in Sec.II.

\item By applying PMC scale setting, all non-conformal $\beta$-terms are absorbed into the running coupling such as to set the optimal renormalization scale of the process at each order. For example, in QED all vacuum polarization contributions to the photon propagator, proper and improper, are summed into the running coupling $\alpha(q^2)$ when one identifies the renormalization scale with the photon virtuality in the Gell Mann-Low scheme. This treatment, which is in effect ``$\beta$-resummation", generally leads to a larger pQCD prediction than the conventional choice of scale, as shown by Tables \ref{tab2}, \ref{tab8}, and \ref{tab14}.

\end{itemize}

\begin{table}[htb]
\centering
\begin{tabular}{|c||c|c|c|c|c|c|}
\hline
 & \multicolumn{3}{c|}{Conventional scale setting} & \multicolumn{3}{c|}{PMC scale setting } \\
\hline
~$\mu^{\rm init}_r$~  & ~$m_t/2$~  & ~$m_t$~ & ~$2m_t$~ & ~$m_t/2$~ & ~$m_t$~ & ~$2m_t$~  \\
\hline
$\sigma^{7\rm TeV}_{(\rm tot)}$ & 165.943 & 161.686 & 154.720 & 168.710 & 168.720 & 168.728  \\
\hline
$\sigma^{8\rm TeV}_{(\rm tot)}$ & 236.232 & 230.557 & 220.797 & 240.337 & 240.351 & 240.362  \\
\hline
$\sigma^{14\rm TeV}_{(\rm tot)}$ & 909.411 & 893.006 & 857.735 & 927.312 & 927.356 & 927.391  \\
\hline
\end{tabular}
\caption{The SM predictions for the top-pair production cross sections (in unit: pb) assuming conventional versus PMC scale settings at the LHC, where $\mu_{r}^{\rm init}=m_t/2$, $m_t$, and $2m_t$, respectively. The factorization scale is taken as $\mu_{f}=m_t$. } \label{LHCpmco}
\end{table}

To discuss the renormalization scale dependence, we present the top-pair cross sections before and after PMC scale setting in Table \ref{LHCpmco};  the contributions from the four production channels are included. Three CM  collision energies $\sqrt{S}=7$ TeV, $8$ TeV, and 14 TeV, and three typical choices of initial scale $\mu_{r}^{\rm init}=m_t/2$, $m_t$, and $2m_t$ have been assumed. The top-pair production cross sections using conventional scale setting show large dependences on the renormalization scale; e.g. the total cross sections for $\mu_{r}^{\rm init}\in [m_t/2,2m_t]$ are $\sigma^{7\rm TeV}_{(\rm tot)}= 162^{+4}_{-7}$ pb, $\sigma^{8\rm TeV}_{(\rm tot)}= 231^{+6}_{-9}$ pb, and $\sigma^{14\rm TeV}_{(\rm tot)}= 893^{+16}_{-35}$ pb, respectively.

Table \ref{LHCpmco} shows that after applying the PMC, the renormalization scale uncertainty can be eliminated even at the NNLO level. The PMC predictions for the total cross sections are very close to $\sigma^{7\rm TeV}_{(\rm tot)}\simeq 169$ pb, $\sigma^{8\rm TeV}_{(\rm tot)}\simeq 240$ pb, and $\sigma^{14\rm TeV}_{(\rm tot)}\simeq 927$ pb. There is a residual scale dependence due to unknown higher-order $\{\beta_i\}$-terms, which is highly suppressed~\cite{pmc1,pmc2,pmc4,pmc6,pmc8,pmc9,pmc11}. These PMC predictions are in excellent agreement with the CMS and ATLAS measurements with $\sqrt{S}=7$ TeV~\cite{CMStot0,CMStot1,CMStot2,ATLtot0,ATLtot1} and $\sqrt{S}=8$ TeV~\cite{CMStot81,CMStot82,ATLAStot81,ATLAStot82}.

\subsection{The top-quark charge asymmetry at the LHC}

As discussed in the above subsection, the renormalization scale dependence for the total cross sections can be eliminated by applying the PMC. We will now show how the top-quark charge asymmetry is affected. For this purpose, we adopt Eq.(\ref{pmcasy}) to do the PMC calculations. As for the numerical results, if not specially stated, we shall always take $\mu_f=m_t$.

\begin{figure}[htb]
\includegraphics[width=\linewidth]{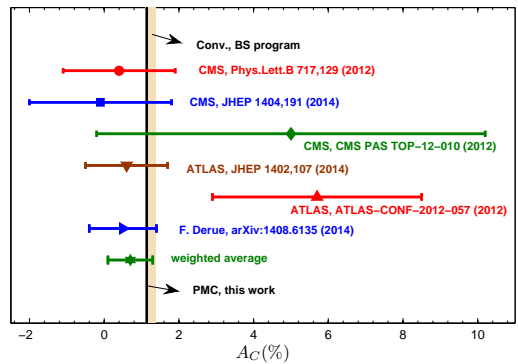}
\caption{The top-quark charge asymmetry $A_{\rm C}$ assuming conventional scale setting (Conv.) and PMC scale setting for $\sqrt{S}=7$ TeV; the error bars are for $\mu^{\rm init}_r \in[m_t/2,2m_t]$ and $\mu_f \in[m_t/2,2m_t]$. As a comparison, the experimental results~\cite{CMS1,CMS2,CMS3,ATLAS1,ATLAS2,ATLAS3} are also presented. }
\label{cpmcasy}
\end{figure}

At the LHC with $\sqrt{S}=7$ TeV, we have
\begin{eqnarray}
\sigma^{\rm BS,LO}_{\rm tot} &=& 100.076 ~{\rm pb}, ~\left. A^{\rm (BS)}_{\rm C}\right|_{\alpha^3_s}=1.068 ~\%, \nonumber \\
\left. A^{\rm (BS)}_{\rm C}\right|_{\alpha^2_s\alpha} &=& 0.124~\%, ~\left. A^{\rm (BS)}_{\rm C}\right|_{\alpha^2}=0.039 ~\%, \nonumber
\end{eqnarray}
which includes all the asymmetric $q\bar{q}$, $gq$ and $g\bar{q}$ channels' contributions. After applying the PMC, we obtain
\begin{eqnarray}
\sigma^{\rm PMC,NLO}_{\rm tot} = 157.596 ~{\rm pb}. \nonumber
\end{eqnarray}
Following the idea of Ref.\cite{pmc3}, the effective coupling constant is
\begin{eqnarray}
{\overline{\alpha}_s} \left(\overline{\mu}^{\rm PMC, NLO}_r \right) = 0.1233; \nonumber
\end{eqnarray}
thus $\overline{\mu}^{\rm PMC, NLO}_r \sim 69$ GeV. The results for the top-quark charge asymmetry are presented in Fig.(\ref{cpmcasy}) in which the CMS and ATLAS measurements~\cite{CMS1,CMS2,CMS3,ATLAS1,ATLAS2,ATLAS3} are included for comparison. Assuming conventional scale setting, we obtain the charge asymmetry $A_{\rm C}=(1.23\pm0.14)\%$. After applying the PMC, it improves to $\left(1.15^{+0.01}_{-0.03}\right)\%$, where the scale errors assume the ranges $\mu^{\rm init}_r \in[m_t/2,2m_t]$ and $\mu_f \in[m_t/2,2m_t]$.

\begin{figure}
\includegraphics[width=\linewidth]{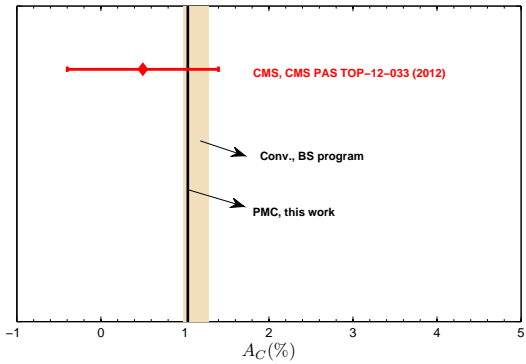}
\caption{The top-quark charge asymmetry $A_{\rm C}$ assuming conventional scale setting (Conv.) and PMC scale setting for $\sqrt{S}=8$ TeV; the error bars are for $\mu^{\rm init}_r \in[m_t/2,2m_t]$ and $\mu_f \in[m_t/2,2m_t]$. The CMS measurement~\cite{CMS8T} is also presented.}
\label{c8pmcasy}
\end{figure}

At the LHC with $\sqrt{S}=8$ TeV, we have
\begin{eqnarray}
\sigma^{\rm BS,LO}_{\rm tot} &=& 142.656 ~{\rm pb}, ~\left. A^{\rm (BS)}_{\rm C}\right|_{\alpha^3_s}=0.960 ~\%, \nonumber \\
\left. A^{\rm (BS)}_{\rm C}\right|_{\alpha^2_s\alpha} &=& 0.110~\%, ~\left. A^{\rm (BS)}_{\rm C}\right|_{\alpha^2}=0.035 ~\%. \nonumber \\
\sigma^{\rm PMC,NLO}_{\rm tot} &=& 225.046 ~{\rm pb},
~{\overline{\alpha}_s} \left(\overline{\mu}^{\rm PMC, NLO}_r \right) = 0.1233. \nonumber
\end{eqnarray}
The results for the top-quark charge asymmetry are presented in Fig.(\ref{c8pmcasy}), in which the CMS measurement~\cite{CMS8T} is also presented as a comparison. Assuming conventional scale setting, we obtain the charge asymmetry $A_{\rm C}=\left(1.11^{+0.17}_{-0.13}\right)\%$. After applying the PMC, it improves to $\left(1.03^{+0.01}_{+0.00}\right)\%$, where the scale errors are for $\mu^{\rm init}_r \in[m_t/2,2m_t]$ and $\mu_f \in[m_t/2,2m_t]$.

\begin{figure}
\includegraphics[width=\linewidth]{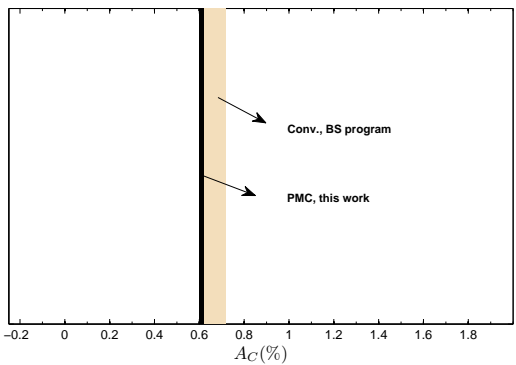}
\caption{The top-quark charge asymmetry $A_{\rm C}$ assuming conventional scale setting (Conv.) and the PMC scale setting for $\sqrt{S}=14$ TeV; the error bars are for $\mu^{\rm init}_r \in[m_t/2,2m_t]$ and $\mu_f\in[m_t/2,2m_t]$. } \label{c14pmcasy}
\end{figure}

At the LHC with $\sqrt{S}=14$ TeV, we have
\begin{eqnarray}
\sigma^{\rm BS,LO}_{\rm tot} &=& 551.721 ~{\rm pb}, ~\left. A^{\rm (BS)}_{\rm C}\right|_{\alpha^3_s}=0.575 ~\%, \nonumber\\
\left. A^{\rm (BS)}_{\rm C}\right|_{\alpha^2_s\alpha} &=& 0.072~\%, ~\left. A^{\rm (BS)}_{\rm C}\right|_{\alpha^2}=0.022 ~\%, \nonumber\\
\sigma^{\rm PMC,NLO}_{\rm tot} &=& 874.849 ~{\rm pb},
~{\overline{\alpha}_s} \left(\overline{\mu}^{\rm PMC, NLO}_r \right) = 0.1233. \nonumber
\end{eqnarray}
The results for the top-quark charge asymmetry for $\sqrt{S}=14$ TeV are presented in Fig.(\ref{c14pmcasy}). Assuming conventional scale setting, we obtain the charge asymmetry $A_{\rm C}=(0.67\pm0.05)\%$. After applying the PMC, it improves to $\left(0.62^{+0.00}_{-0.02}\right)\%$, where the scale errors are for $\mu^{\rm init}_r \in[m_t/2,2m_t]$ and $\mu_f \in[m_t/2,2m_t]$.

\begin{figure}[tb]
\includegraphics[width=0.45\textwidth]{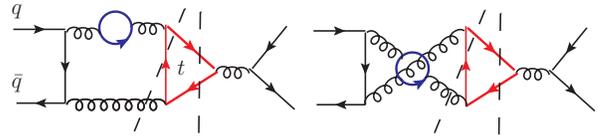}
\caption{Cut diagrams for the $n_f$-terms at the $\alpha^4_s$-order of the asymmetric $(q\bar{q})$-channel; these cuts lead to a small effective NLO PMC scale $\overline{\mu}^{\rm PMC, NLO}_r$, where the solid circles stand for the light quark loops. }
\label{nloscale}
\end{figure}

The charge asymmetry decreases with increasing CM collision energy $\sqrt{S}$. This is reasonable, since the asymmetry is diluted by the symmetric $(gg)$-channel at the LHC, and the ratio between the subprocess cross sections for the $(gg)$ the $(q\bar{q})$, and the $\sigma_{gg}/\sigma_{q\bar{q}}$ channels are equal to $4.8,5.5,9.2$ for $\sqrt{S}=7,8,14$ TeV, respectively. We observe that the effective PMC scale, $\overline{\mu}^{\rm PMC, NLO}_r|_{\rm LHC} \sim 69$ GeV, is the same for all three cases. This shows that the effective renormalization scale is independent of the collision energy, since the running behavior for the strong coupling constant is determined by its $\beta$-function. As explained in Ref.\cite{pmc3}, the small effective PMC scale is dominated by the non-Coulomb $n_f$-terms of the dominant asymmetric $(q\bar{q})$-channel at the $\alpha^4_s$-order, which are shown in Fig.(\ref{nloscale}). The PMC scale $\overline{\mu}^{\rm PMC, NLO}_r$ is a weighted average of the different momentum flows within the gluons; it thus can be small.

In summary, Figs.(\ref{cpmcasy},\ref{c8pmcasy},\ref{c14pmcasy}) indicate that after applying the PMC, the renormalization scale uncertainty can be greatly suppressed and shows better agreement with the present CMS and ATLAS data.\\

\subsection{The top-quark charge asymmetry at the LHC for the  kinematic cut $M_{t\bar{t}}>M_{\rm cut}$}

In order to compare with future data at the LHC, it will be useful to calculate the dependence of $A_{\rm C}$ on the $t\bar{t}$-invariant mass $M_{t\bar{t}}$, $A_{\rm C}(M_{t\bar{t}}>M_{\rm cut})$.

\begin{widetext}
\begin{center}
\begin{table}[htb]
\begin{tabular}{|c||c|c|c|c|c|c|c|c|c|}
\hline
& \multicolumn{3}{c|}{7 TeV $(A_{\rm C}(M_{t\bar{t}}>M_{\rm cut}))$} & \multicolumn{3}{c|}{8 TeV $(A_{\rm C}(M_{t\bar{t}}>M_{\rm cut}))$} & \multicolumn{3}{c|}{14 TeV $(A_{\rm C}(M_{t\bar{t}}>M_{\rm cut}))$} \\
\hline
  ~$M_{\rm cut}$~ & ~0.5 TeV~ & ~0.7 TeV~ & ~1 TeV~ & ~0.5 TeV~ & ~0.7 TeV~ & ~1 TeV~ & ~0.5 TeV~ & ~0.7 TeV~ & ~1 TeV~ \\
\hline\hline
~Conv.~\cite{wsAfb}~ & ~1.48\%~ & ~1.95\%~ & ~2.46\%~ & ~1.40\%~ & ~1.84\%~ & ~2.32\%~ & ~0.86\%~ & ~0.98\%~ & ~1.34\%~ \\
\hline
~PMC~ & ~2.67\%~ & ~1.65\%~ & ~0.99\%~ & ~2.39\%~ & ~1.51\%~ & ~0.93\%~ & ~1.28\%~ & ~0.74\%~ & ~0.51\%~ \\
\hline\hline
~$R_{C}$~ & $0.78$ & $0.15$ & $0.60$ & $0.71$ & $0.18$ & $0.60$ & $0.49$ & $0.25$ & $0.62$ \\
\hline
\end{tabular}
\caption{Top-pair charge asymmetries $A_{\rm C}(M_{t\bar{t}}>M_{\rm cut})$, taking various kinematic cuts for the $t\bar{t}$-invariant mass $M_{t\bar{t}}$ at the LHC and the CM collision energy $\sqrt{S}=7$ TeV, 8 TeV, and 14 TeV, respectively. The results for conventional scale setting (Conv.) and PMC scale setting are presented. The values of the ratio $R_{C}(M_{t\bar{t}}>M_{\rm cut})$ are also presented. $\mu^{\rm init}_r=\mu_f=m_t$. } \label{tabACMtt}
\end{table}
\end{center}
\end{widetext}

The top-pair asymmetries at the LHC for several typical cuts $M_{\rm cut}=0.5$ TeV, 0.7 TeV, and 1 TeV, are presented in Table \ref{tabACMtt}. In order to show how the charge asymmetries change before and after PMC scale setting, we define the ratio,
\begin{displaymath}
R_{C} = \left|\frac{A_{\rm C}(M_{t\bar{t}}>M_{\rm cut})|_{\rm PMC}- A_{\rm C}(M_{t\bar{t}}>M_{\rm cut})|_{\rm Conv.}} {A_{\rm C}(M_{t\bar{t}}>M_{\rm cut})|_{\rm Conv.}}\right|.
\end{displaymath}
In the case of $M_{\rm cut}=0.5$ TeV, the value of $R_{C}$ changes to $0.78$, $0.71$, $0.49$ for $\sqrt{S}=7,8,14$ TeV, respectively. This again demonstrates that the proper choice of renormalization scale is essential. Table \ref{tabACMtt} shows that for a large value of the $t\bar{t}$-invariant mass as $M_{\rm cut} > 0.5$ TeV, $A_{\rm C}$ increases with increasing $M_{\rm cut}$ assuming conventional scale setting, but it decreases with increasing $M_{\rm cut}$ after applying the PMC. This can be qualitatively explained by the following points: I) the cross sections at the LO level and at the NLO level for ($q\bar{q}$)- and ($gg$)-channels rapidly decrease with increasing $M_{\rm cut}$, whereas only small changes are found for the ($gq$) and ($g\bar{q}$)-channels; II) These two channels' relative contributions to $\sigma^{\rm PMC,NLO}_{\rm tot}$ at the NLO level are thus increased in comparison to their contributions at the LO level; III) the effective NLO PMC scale $\overline{\mu}^{\rm PMC, NLO}_r$ increases with increasing $M_{\rm cut}$; i.e. for the case of $\sqrt{S}=7$ TeV, we have $\overline{\mu}^{\rm PMC, NLO}_r \sim 30$ GeV, $92$ GeV, and $143$ GeV for $M_{\rm cut}=0.5$ TeV, $0.7$ TeV, and $1.0$ TeV, respectively. Then, by using Eq.(\ref{pmcasy}), we obtain a decreasing asymmetry $A_{\rm C}$ with increasing  $M_{\rm cut}$.

\subsection{An estimate of the factorization scale dependence of the top-quark charge asymmetry}

As seen in Table \ref{LHCpmco}, the dependence on the choice of the initial renormalization scale is greatly suppressed after applying PMC scale setting. The remaining dominant errors are from the factorization scale dependence.  The determination of the factorization scale is a completely separate issue from the renormalization scale setting since it is present even for a conformal theory with $\beta=0$. The factorization scale should be chosen to match the nonpertubative bound-state dynamics with perturbative DGLAP evolution~\cite{dglap1,dglap2,dglap3}. This can be done explicitly by using  nonperturbative models such as the AdS/QCD and the light-front holography where the light-front wavefunctions of the hadrons are known~\cite{adsQCD}.

Fortunately, we find that the factorization scale dependence is suppressed after applying the PMC; this can be explained by the fact that the pQCD series behaves much better after applying the PMC. To show clearly how the choice of  factorization scale affects the asymmetry, we fix the initial renormalization scale $\mu^{\rm init}_r =m_t$.

Using conventional scale setting, we obtain
\begin{eqnarray}
A_{C|7\rm TeV} &=& \left(1.23^{+0.04}_{-0.05}\right)\%, \nonumber\\
A_{C|8\rm TeV} &=& \left(1.11^{+0.08}_{-0.04}\right)\%, \nonumber\\
A_{C|14\rm TeV}&=& \left(0.67^{-0.00}_{-0.01}\right)\%.\nonumber
\end{eqnarray}
After applying  PMC scale setting, we obtain
\begin{eqnarray}
A_{C|7\rm TeV} &=& \left(1.15^{+0.01}_{-0.03}\right)\%,  \nonumber \\
A_{C|8\rm TeV} &=& \left(1.03^{+0.01}_{+0.00}\right)\%,  \nonumber \\
A_{C|14\rm TeV}&=& \left(0.62^{+0.00}_{-0.02}\right)\%.  \nonumber
\end{eqnarray}
Here, the central values are for $\mu_f=m_t$, and the errors are for $\mu_{f} \in[m_t/2,2m_t]$. It is obvious that the factorization scale dependence is decreased after applying the PMC. As an explanation, we can re-express the log-terms of the form $\ln^{k}(\mu_r^2/\mu_f^2)$ as $\left(\ln\mu_r^2/m_t^2 - \ln\mu_f^{2}/m_t^{2}\right)^{k}$. Because of the correlation of $\ln^{m}\mu_r^2/m_t^2$ and $\ln^{m}\mu_f^2/m_t^2$, the simple conventional scale-setting procedure of setting $\mu_r=m_t$ to eliminate the log-terms $\ln^{k}\mu_r^2/m_t^2$ is again problematic, since it may lead to a large factorization scale dependence. This again explains the importance of proper renormalization scale setting.

\section{Summary}  \label{sect4}

In the present paper, we have made a detailed comparison of the top-quark charge asymmetry at the LHC before and after PMC scale setting.

The setting of the renormalization scale of the QCD coupling is one of the outstanding fundamental problems of pQCD. The elimination of this systematic error is essential for precision tests of theory at colliders such as the LHC and for increasing the sensitivity of experiment to new physics. The PMC provides a systematic and unambiguous procedure to set the renormalization scale for any QCD process at any finite order of perturbation theory. The PMC predictions are also scheme independent as required by renormalization group invariance.

As shown in Tables \ref{tab2}, \ref{tab8}, and \ref{tab14}, we do achieve a pQCD series with improved convergence for the top-pair production cross sections at the LHC up to NNLO level after applying the PMC. Taking the dominant asymmetric $(q\bar{q})$-channel as an example, one obtains $|\sigma_{q\bar{q}}^{\rm NNLO}/ \sigma_{q\bar{q}}^{\rm NLO}|_{\rm 7 TeV} \sim 54\%$ for conventional scale setting; it reduces to $\sim 10\%$ after applying PMC scale setting.

As shown in Table \ref{LHCpmco}, the conventional renormalization scale uncertainty for the top-pair productions up to NNLO level has been almost eliminated by the PMC. The PMC predictions for the total cross section are essentially fixed to $\sigma^{7\rm TeV}_{(\rm tot)}\simeq 169$ pb, $\sigma^{8\rm TeV}_{(\rm tot)}\simeq 240$ pb, and $\sigma^{14\rm TeV}_{(\rm tot)}\simeq 927$ pb. The PMC predicts that the effective momentum flow for the top-pair production using the $\overline{\rm MS}$ scheme is close to $m_t/2$, far from the guessed value of $m_t$, which is determined from the naive idea of eliminating the large log terms as $\ln^k m_t^2/\mu_r^2$.

\begin{table}[htb]
\begin{tabular}{|c|c|c|c|}
\hline
   & ~7 TeV~ & ~8 TeV~ & ~14 TeV~  \\
\hline
~Conv.~ & $\left(1.23^{+0.14}_{-0.14}\right)\%$~ & ~$\left(1.11^{+0.17}_{-0.13}\right)\%$~ & ~$\left(0.67^{+0.05}_{-0.05}\right)\%$~  \\
\hline
~PMC~  & ~$\left(1.15^{+0.01}_{-0.03}\right)\%$~ & ~$\left(1.03^{+0.01}_{+0.00}\right)\%$~ & ~$\left(0.62^{+0.00}_{-0.02}\right)\%$~ \\
\hline
\end{tabular}
\caption{The top quark charge asymmetries assuming conventional scale setting versus PMC scale setting at the LHC with $\sqrt{S}=7$ TeV, 8 TeV, and 14 TeV, respectively. The results for the conventional scale setting (Conv.) and the PMC scale setting are presented. $\mu^{\rm init}_r \in[m_t/2,2m_t]$ and $\mu_f\in[m_t/2,2m_t]$. } \label{tab3}
\end{table}

We summarize the top quark charge asymmetries before and after PMC scale setting in Table \ref{tab3}. After applying PMC scale setting, the asymmetries have much smaller scale dependence; the resulting predictions are also in better agreement with the available ATLAS and CMS data within errors. At present the LHC measurements still have large errors; the comparison of the predicted asymmetry with future more accurate data will improve our understandings on the top-quark asymmetry.

It is often argued that by varying the renormalization scale within a certain region, one can predict unknown higher-order contributions. However such scale variation can only predict part of the higher-order contributions  -- the $\beta$ terms -- and it is scheme dependent~\cite{pmc8}. On the other hand, the PMC series is identical to the corresponding conformal series; the divergent renormalon terms are absent. After PMC scale setting, the scales are optimized  at each order of perturbation theory and cannot be varied; otherwise, one explicitly breaks the renormalization group invariance, which then leads to an unreliable prediction. A conservative way to predict the unknown higher-order contributions after PMC scale setting has been suggested in Ref.\cite{pmc8}. A detailed comparison of the predicted unknown higher-orders contributions using various renormalization scale settings has been given in ref.~\cite{pmc8}. Two observables $R(e^+e^-)$ and $\Gamma(H\to b\bar{b})$ are studied up to four-loop level~\cite{pmc8}. The figures (4,5) in this reference show that the error bars from ``unknown" higher-order corrections quickly approach a stable values when one applies PMC scale setting in strong contrast to conventional scale setting.  The values for the ``unknown" higher-order terms are well within the error bars predicted from the corresponding one-order lower terms.

We have also calculated the top-quark charge asymmetries assuming several typical invariant top-pair invariant mass cuts. Table \ref{tabACMtt} shows that after applying the PMC, the asymmetry $A_{\rm C}$ decreases with increasing  $M_{\rm cut}$. The difference in predicted asymmetries between PMC and convention scale setting again shows that the proper choice of  renormalization scale is essential.

We take this opportunity to emphasize two additional features for the PMC scale setting:
\begin{itemize}
\item A demonstration of the renormalization scheme dependence of pQCD predictions has been done in Refs.\cite{pmc6,pmc11} by introducing  a generalized ${\rm MS}$-like renormalization scheme with an arbitrary subtraction constant $\delta_i$ at each order; i.e., the $R_\delta$-scheme. The $R_\delta$-scheme provides a systematic,  process-independent way to identify the $\beta_i$ terms at each perturbative order. The resulting ``degeneracy relations", achieved by setting $\{\delta_i\}=1$ in Eq.(4) of Ref.\cite{pmc6}, demonstrates that the scheme-independent conformal terms are the same for any ${\rm MS}$-like renormalization scheme. The PMC predictions for physical observables are independent of the choice of renormalization scheme.

\item The PMC provides a process-independent way to absorb all $\beta$-terms into the running coupling, consistent with its renormalization group equation. The $\beta$-terms that determine the running behavior are absorbed into the running coupling to form a new PMC scale (optimal renormalization scale) and effective number of flavors $n_f$ at each specific $\alpha_s$ order, as in the Gell Mann Low procedure in QED. One can confirm that the non-conformal $\beta$-terms are correctly identified and absorbed by the PMC procedure by checking that there is negligible dependence of the fixed-order theory prediction on the initial scale.
\end{itemize}

\noindent{\bf Acknowledgments}: This work was supported in part by the Fundamental Research Funds for the Central Universities under Grant No.CQDXWL-2012-Z002 and by the Natural Science Foundation of China under Grant No.11275280 and No.11325525, the Department of Energy Contract No.DE-AC02-76SF00515, and by the Open Project Program of State Key Laboratory of Theoretical Physics, Institute of Theoretical Physics, Chinese Academy of Sciences under Grant No.Y3KF311CJ1. SLAC-PUB-16116.

\end{document}